\documentclass[11pt]{article}
\usepackage{amsmath,amssymb,color}

\usepackage{amsfonts}

\newcommand{\hoch}[1]{$\, ^{#1}$}

\newcommand{\be}{\begin{equation}}
\newcommand{\ee}{\end{equation}}
\newcommand{\bea}{\setlength\arraycolsep{2pt} \begin{eqnarray}}
\newcommand{\eea}{\end{eqnarray}}

\def\ft#1#2{{\textstyle{\frac{\scriptstyle #1}{\scriptstyle #2} } }}
\def\fft#1#2{{\frac{#1}{#2}}}

\def\0{{\sst{(0)}}}
\def\1{{\sst{(1)}}}
\def\2{{\sst{(2)}}}
\def\3{{\sst{(3)}}}
\def\4{{\sst{(4)}}}
\def\5{{\sst{(5)}}}
\def\6{{\sst{(6)}}}
\def\7{{\sst{(7)}}}
\def\8{{\sst{(8)}}}
\def\sst#1{{\scriptscriptstyle #1}}

\begin{document}

\begin{flushright}
\hfill{KIAS-P12028}
\end{flushright}

\vspace{25pt}
\begin{center}
{\large {\bf Supersymmetric Asymptotic AdS and Lifshitz
Solutions}}

{\large{\bf in Einstein-Weyl and Conformal Supergravities}}

\vspace{10pt}

H. L\"u\hoch{1} and Zhao-Long Wang\hoch{2}

\vspace{10pt}

\hoch{1}{\it Department of Physics, Beijing Normal University,
Beijing 100875, China}

\vspace{10pt}

\hoch{2} {\it School of Physics, Korea Institute for Advanced Study,
Seoul 130-722, Korea}

\vspace{40pt}

\underline{ABSTRACT}
\end{center}

We show that there exist supersymmetric Lifshitz vacua in off-shell
Einstein-Weyl supergravity, in addition to the BPS AdS$_4$ vacuum.
The Lifshitz exponents are determined by the product of the
cosmological constant and the coupling of the Weyl-squared term.  We
then obtain the equations of the supersymmetric solutions that are
asymptotic to the AdS or Lifshitz vacua.  We obtain many examples of
exact solutions as well as numerical ones.  We find examples of
extremal AdS black branes whose near-horizon geometry is
AdS$_2\times T^2$. We also find an extremal Lifshitz black hole with
$z=-2$, whose horizon coincides with the curvature singularity.
However the asymptotic Lifshitz solutions are in general smooth
wormholes. In conformal supergravity, we find intriguing examples of
non-extremal ``BPS'' AdS and Lifshitz black holes whose local
Killing spinor is divergent on the horizon.  We show that all the
supersymmetric asymptotic AdS and Lifshitz solutions have the
vanishing Noether charge associated with some scaling symmetry. We
also study the integrability condition of the Killing spinor
equation and the supersymmetric invariance of the action. Finally we
show that the only spherically-symmetric BPS solution is the AdS
vacuum.

\vfill {\footnotesize Emails: mrhonglu@gmail.com;\ \ \
zlwang@kias.re.kr}

\thispagestyle{empty}





\newpage

\section{Introduction}

One of the most important advances in string theory is the AdS/CFT
correspondence \cite{mald,gkp,wit}, which is a duality between a
conformal field theory and a closed string theory on the AdS
background. The principle of holography underlying the AdS/CFT has
been shown to be applicable for broader classes of gauge/gravity
duality outside the string theory.  For asymptotic AdS backgrounds,
the corresponding field theories are Lorentz invariant. In the
context of condensed matter physics, various systems exhibit a
dynamical scaling near fixed points
\begin{equation}\label{scaling}
t\rightarrow \lambda^z t,\qquad x_i\rightarrow \lambda\, x_i,\qquad
z\neq1\,.
\end{equation}
In other words, rather than obeying the conformal scale invariance
$t\rightarrow \lambda\, t$, $x_i\rightarrow \lambda\, x_i$, the
temporal and the spatial coordinates scale anisotropically.

Requiring also time and space translation invariance, spatial
rotational symmetry, spatial parity and time reversal invariance,
the authors of \cite{Kachru:2008yh} were led to consider $D$
dimensional geometries of the form
\begin{equation}\label{L-metric}
    ds^2=\ell^2\biggl(-r^{2z}dt^2+r^2dx^idx^i+\frac{dr^2}{r^2}\biggr)\,.
\end{equation}
(See also \cite{Koroteev:2007yp}.) This metric obeys the scaling
relation (\ref{scaling}) if one also scales
$r\rightarrow\lambda^{-1}r$. If $z=1$, the metric reduces to the
usual AdS metric in Poincar\'e coordinates with AdS radius $\ell$.
Metrics of the form (\ref{L-metric}) can be obtained as solutions in
general relativity with a negative cosmological constant if
appropriate matter is included. For example, solutions were found by
introducing 1-form and 2-form gauge fields \cite{Kachru:2008yh}; a
massive vector field \cite{Taylor:2008tg}; in an abelian Higgs model
\cite{Gubser:2009cg}; and with a charged perfect fluid
\cite{Hartnoll:2010gu}. A class of Lifshitz black hole solutions
with non-planar horizons was found in
\cite{Danielsson:2009gi,Mann:2009yx}.  The string and supergravity
embedding has been found in \cite{Hartnoll}-\cite{Amado:2011nd}.

  Whilst supersymmetric AdS vacua and asymptotic AdS solutions are
plenty and can be simple in string and supergravities, few examples
of supersymmetric Lifshitz and asymptotic Lifshitz solutions have
been constructed and they all involve some complicated
hypermultiplets in addition to the much simpler vector multiplet.
Although the advantage of supersymmetry in condensed matter system
is less evident, the supersymmetry helps to stabilize classical
solutions and enables us to check the AdS/CFT correspondence to the
higher order. However, the complexity of these BPS Lifshitz
solutions makes the task difficult.

   Recently, it was shown that Lifshitz vacua and Lifshitz black
holes arise naturally in higher-derivative pure gravities
\cite{lppp}. (See also \cite{sol1,sol2,sol3} for recent construction
of exact solutions in extended gravities with quadratic curvature
terms.) This is related to the fact that higher-order gravities
contain massive spin-2 modes as well as the massless graviton.  The
massive spin-2 mode plays an analogous role of massive vector in
supporting the Lifshitz black hole.

    Extended gravities with quadratic curvature terms can be
supersymmetrized in the off-shell formalism using a chiral
superfield \cite{Binetruy:2000zx}.  In particular, the Einstein-Weyl
supergravity was studied in detail recently \cite{lpsw}.  The theory
admits a supersymmetric AdS vacuum, and its linear spectrum was
obtained. In this paper, we demonstrate that in addition to the AdS
vacuum, the Einstein-Weyl supergravity admits BPS Lifshitz solutions
that preserve $\fft14$ of the supersymmetry.  The Lifshitz exponent
$z$ is determined by the product of the cosmological constant and
the coupling of the Weyl-squared term.

      The paper is organized as follows.  In section 2, we review
Einstein-Weyl supergravity.  The bosonic field content of the theory
consists of the metric as well as a massive vector and a complex
scalar.  For special choice of parameters, the theory becomes
conformal supergravity.  In section 3, we obtain Lifshitz solutions.
There are two types of such solutions.  The first of which were
obtained previously in Einstein-Weyl pure gravity \cite{lppp} and
the solutions are not supersymmetric.  The second type of solutions
are new with the massive vector field turned on.  We verify by an
explicit construction of Killing spinors that these solutions
preserve $\fft14$ of the supersymmetry.  In section 4, we obtain the
equations governing the supersymmetric backgrounds that are
asymptotic to the AdS or the Lifshitz vacua. We obtain many examples
of exact solutions in Einstein-Weyl supergravity and conformal
supergravity. We then use numerical analysis to discuss the general
feature of these solutions.  All the asymptotic Lifshitz solutions
are plane-symmetric.  Such a solution has a new global scaling
symmetry and the associated Noether charge is a conserved quantity.
We find that owing to the supersymmetry, the BPS solutions we have
obtained all have the vanishing Noether charge. We present this
analysis in section 5.

      An important feature of off-shell supergravity is that one can
add additional super invariants to the action without modifying the
supersymmetric transformation rules.  This makes the supersymmetric
invariance of the action of off-shell supergravity very different
from that of usual on-shell supergravity. In section 6, we study the
integrability condition of the Killing spinor equation, and also the
invariance of the action.  Some new features arise in off-shell
supergravity.  In section 7, we consider spherically-symmetric
solutions. We demonstrate that only supersymmetric such a solution
in Einstein-Weyl supergravity is the AdS vacuum. We conclude the
paper in section 8.

\section{Einstein-Weyl supergravity}

The field content of the off-shell ${\cal N}=1$, $D=4$ supergravity
consists of the metric $e_{\mu}^a$, a massive vector $A_\1$ and a
complex scalar $S + {\rm i} P$, totalling 12 off-shell degrees of
freedom, matching with that of the gravitino $\psi_\mu$.  The
general formalism for constructing a supersymmetric action for any
chiral superfield was obtained in \cite{Binetruy:2000zx}.  For
appropriate choices of superfields, one obtains the actions of the
supersymmetrizations of the cosmological term, the Einstein Hilbert
term and higher-order curvature terms.  Excluding the higher-order
curvature terms, the massive vector and scalars are both auxiliary
and are set to zero by the equations of motion, leading to ${\cal
N}=1$ supergravity in four dimensions.  However, when higher-order
curvature invariants are included, these auxiliary fields may
develop dynamics and cannot be truncated out. At the quadratic order
of the curvature, there are two super invariants and one of them is
the Weyl-squared invariant. In this paper we shall consider
including only this quadratic invariant, which has a consequence
that the vector field becomes dynamical whilst the complex scalar
remains auxiliary. (The other quadratic invariant will turn on the
complex scalar as well.) Adopting the notation of \cite{lpsw}, the
bosonic Lagrangian is given by
\begin{equation}
e^{-1}{\cal L} = R - \ft23 (A_\1^2 + S^2 + P^2) + 4S\,
\sqrt{-\Lambda/3} + \ft12\alpha C^{\mu\nu\rho\sigma}
C_{\mu\nu\rho\sigma} + \ft13\alpha F_\2^2\,,\label{genlag}
\end{equation}
where $C$ is the Weyl tensor and $F_\2=dA_\1$.  Note that the
quantity $m=\sqrt{-\Lambda/3}$ can take both positive and negative
values. The supersymmetric transformation rule for the gravitino is
given by
\begin{equation}
\delta \psi_\mu = -D_{\mu} \epsilon + \ft{1}6 (2A_\mu -
\Gamma_{\mu\nu} A^\nu)\Gamma_5 \epsilon - \ft16\Gamma_\mu (S + {\rm
i}\Gamma_5 P)\epsilon\,.\label{susytrans}
\end{equation}
The full supersymmetric action and transformation rules will be
discussed in section 6. The leading-order supergravity with
$\alpha=0$ (and also $\Lambda=0$) was constructed much earlier in
\cite{sw,fv}. Note that compared to \cite{lpsw}, we have sent vector
field $A_\1$ to ${\rm i} A_\1$ in (\ref{genlag}) and
(\ref{susytrans}), which can be clearly done for auxiliary fields in
off-shell supergravities. This is because the vector field, being
part of the auxiliary multiplet, is not associated with any central
charge in the super algebra, and hence sending $A_\1$ to ${\rm i}
A_\1$ does not upset the positiveness of the anti-commutator of the
supercharges.

The equations of motion for the scalar fields $S$ and $P$ imply that
\begin{equation}
S=3\sqrt{-\Lambda/3}\,,\qquad P=0\,,\label{scalareom}
\end{equation}
and hence they are auxiliary with no dynamical degree of freedom.
The equations of motion for the vector field and the metric are given by
\begin{eqnarray}
0&=&\alpha\,\nabla^{\mu} F_{\mu\nu} + A_\nu\,,\cr 
0&=& R_{\mu\nu}- \ft12 R g_{\mu\nu} + \Lambda\, g_{\mu\nu} -
\ft23\alpha
(F_{\mu\nu}^2 - \ft14 F^2 g_{\mu\nu})\cr %
&&+\ft23 (A_\mu A_\nu - \ft12 A^2 g_{\mu\nu}) -\alpha
(2\nabla^\rho\nabla^\sigma + R^{\rho\sigma})C_{\mu\rho\sigma\nu}\,.
\end{eqnarray}
It was shown in \cite{lpsw} that the theory admits a supersymmetric
AdS vacuum with the cosmological constant $\Lambda$.  The linear
spectrum on the AdS background was analyzed.  The gravity modes are
identical to those in Einstein-Weyl gravity studied in
\cite{lpcritical}. (See also
\cite{linear1,linear2,linear3,linear4}.) There is a ghost massive
spin-2 mode in additional to the massless graviton.  The mass is
determined by the quantity $\alpha \Lambda$, the product of the
cosmological constant and the coupling of the Weyl-squared term.
There is a critical phenomenon when
\begin{equation}
\alpha \Lambda = \ft32\,,
\end{equation}
for which the massive spin-2 mode disappears and is replaced by the
log mode \cite{lpcritical}.  It was shown in \cite{lppp} that there
is no Lifshitz solution in critical Einstein-Weyl pure gravity.  As
we shall demonstrate presently, critical Einstein-Weyl supergravity
can admit a $z=4$ supersymmetric Lifshitz solution owing to the
existence of the massive vector field $A_\1$.

    It is worth remarking that owing to
the fact that the theory is off-shell, the supersymmetric
transformation rule (\ref{susytrans}) is independent of the
parameter $\alpha$, even though the equations of motion are clearly
modified by the $\alpha$ term. This in particular means that the
$\alpha$ terms alone, which form conformal gravity, are
supersymmetric.  The full bosonic Lagrangian of off-shell conformal
supergravity is given by
\begin{equation}
e^{-1}{\cal L} =  \ft12\alpha C^{\mu\nu\rho\sigma}
C_{\mu\nu\rho\sigma} + \ft13\alpha F_\2^2\,,\label{conflag}
\end{equation}
with the supersymmetric transformation given by (\ref{susytrans}).
At the first sight, it appears that the conformal symmetry is broken
by the fermions.  However, in this case the complex scalar $S+{\rm
i} P$ couples only to the fermion and there are no bosonic equations
of motion to constrain its value.  Thus, they can be used to restore
the conformal symmetry in the fermionic sector.  Also note that in
the bosonic sector, the vector field becomes massless and hence the
Lagrangian (\ref{conflag}) is gauge invariant. However, it is easy
to see from (\ref{susytrans}) that the gauge invariance is broken by
the fermionic sector.  Thus a supersymmetric solution requires a
specific gauge choice.

\section{Supersymmetric Lifshitz and AdS vacua}

In \cite{lpsw}, it was shown that off-shell Einstein-Weyl
supergravity admits an AdS vacuum, and its linear spectrum was also
discussed. In a recent paper \cite{lppp}, it was shown that
Einstein-Weyl pure gravity admits Lifshitz solutions.  In this
section, we construct supersymmetric Lifshitz vacua.  We consider
the following ansatz
\begin{equation}
ds^2 = \ell^2 \Big(-r^{2z} dt^2 + r^2 (dx^2 + dy^2) +
\fft{dr^2}{r^2}\Big)\,,\qquad A_\1=q r^z dt\,.\label{lifs-ans}
\end{equation}
It is clear the vector field preserves the Lifshitz symmetry. The
equations of motion for the massive vector field $A_\1$ imply that
\begin{equation}
q(1+ \fft{2\alpha z}{\ell^2})=0\,.
\end{equation}
Thus there is a bifurcation of solutions
\begin{equation}
q=0\,,\qquad {\rm or}\qquad \alpha=-\fft{\ell^2}{2z}\,.
\end{equation}
For $q=0$, we recover the Lifshitz solutions constructed in
\cite{lppp}, given by
\begin{equation}
\alpha = \fft{3\ell^2}{2z(z-4)}\,,\qquad \Lambda = -
\fft{3+2z+z^2}{2\ell^2}\,.\label{ewlif}
\end{equation}
Alternatively, for non-vanishing $q$, we have
\begin{equation}
\alpha = -\fft{\ell^2}{2z}\,,\qquad
\Lambda=-\fft{(2+z)^2}{3\ell^2}\,,\qquad q = z-1\,,\qquad
S=\fft{z+2}{\ell}\,.\label{susylif}
\end{equation}
Note that when $z=1$, the solutions of both cases reduce to the
AdS$_4$. However, the AdS$_4$ vacuum is a solution independent of
the value of $\alpha$.

We now examine the supersymmetry of the Lifshitz solution
(\ref{susylif}) by solving for the Killing spinors of $\delta
\psi_\mu=0$ in this background. First let us choose a convenient
choice for the vielbein
\begin{equation}
e^{\hat 0}=\ell\, r^z dt\,,\qquad e^{\hat x} = \ell\, r
dx\,,\qquad e^{\hat y} =\ell\, r dy\,,\qquad e^{\hat
r}=\ell\,\fft{dr}{r}\,,
\end{equation}
where we use the hatted indices to denote those of tangent
space. The non-vanishing components of the corresponding spin
connection are then given by
\begin{equation}
\omega^{\hat 0}{}_{\hat r}= \ell^{-1}\, z e^{\hat
0}\,,\qquad \omega^{\hat x}{}_{\hat r}= \ell^{-1}
e^{\hat x}\,,\qquad \omega^{\hat y}{}_{\hat r}=
\ell^{-1}\, e^{\hat y}\,.
\end{equation}
The Killing spinors satisfy
\begin{eqnarray}
\mu=0:&& \Big(-\ft12 z \Gamma_r + \ft13 (z-1)\Gamma^{\hat
0}\Gamma_5 - \ft16 (z+2)\Big)\epsilon=0\,,\cr 
\mu=x,y:&& \Big(-\ft12 \Gamma_r - \ft16 (z-1)\Gamma^{\hat
0}\Gamma_5 - \ft16 (z+2)\Big)\epsilon=0\,,\cr 
\mu=r:&&\Big(-\partial_r + \ft{z-1}{6r} \Gamma_{\hat r}
\Gamma_5\Gamma^{\hat 0} - \ft{z+2}{6r} \Gamma_{\hat r}
\Big)\epsilon=0\,.
\end{eqnarray}
A solution exists and it is given by
\begin{equation}
\epsilon=\sqrt r\,\epsilon_0\,,\qquad \Gamma_{\hat r}
\epsilon_0+\epsilon_0=0\,,\qquad \Gamma^{\hat
0}\Gamma_5\epsilon_0 +\epsilon_0=0\,,
\end{equation}
Thus we see that there are two independent and commuting projections
on the constant spinors $\epsilon_0$, yielding the conclusion that
the Lifshitz solutions preserve $\ft14$ of the supersymmetry. Note
that for $z=1$, there is a supersymmetric enhancement since the
solutions becomes AdS$_4$, which also admits additional Killing
spinors that depend on $t,x,y$ coordinates \cite{lptks}.

In the limit of $\alpha\rightarrow \infty$, we have $z=0$.  The
solution can be viewed as that of conformal supergravity.  The
solution is somewhat trivial since the vector becomes massless and
pure gauge.  The metric is a direct product of time and a hyperbolic
3-space.

The analogous demonstration can be used to show
that the Lifshitz solution (\ref{ewlif}) of Einstein-Weyl pure
gravity with the vanishing massive vector has no Killing spinor. This
is because the metric and hence the spin connection is of the same
form as that with $A_\1$ turned on. It follows that the contribution
in the Killing spinor equations due to the spin connection is linear
to $z$, and hence it cannot balance the contribution from the
cosmological term which is square root of a quadratic function of
$z$.

      Thus we see that there is no smooth deformation that connects
to the supersymmetric (\ref{susylif}) and non-supersymmetric
(\ref{ewlif}) Lifshitz solutions in Einstein-Weyl supergravity. This
is related the fact that the coupling for the massive vector is
associated with the coupling of the Weyl-squared term by the
supersymmetry. If we relax this condition and consider the following
Lagrangian:
\begin{equation}
e^{-1}{\cal L} = R - \ft23 (A_\1^2 + S^2 + P^2) + 4S\,
\sqrt{-\Lambda/3} + \ft12\alpha C^{\mu\nu\rho\sigma}
C_{\mu\nu\rho\sigma} + \ft13\beta F_\2^2\,,\label{nonsusylag}
\end{equation}
We then have a continuous family of solutions
\begin{equation}
\alpha = - \fft{\ell^2 (3-3z+q^2)}{2z(4-5z+z^2)}\,,\qquad
\beta=-\fft{\ell^2}{2z}\,,\qquad \Lambda = \fft{q^2 - 3
(3+2z+z^2)}{6\ell^2}\,.
\end{equation}
This general solutions contain the ones of Einstein-Weyl pure
gravity (with $q=0$) \cite{lppp} and cosmological Einstein gravity
with a massive vector (with $\alpha=0$) \cite{Taylor:2008tg}.  Of
course, the price for having such a continuous deformation is that
the theory (\ref{nonsusylag}) with $\beta\ne\alpha$ is not
supersymmetric.

    To conclude this section, we remark that the allowed Lifshitz
exponents for both supersymmetric (\ref{susylif}) and
non-supersymmetric (\ref{ewlif}) solutions are determined by the
quantity $\alpha\Lambda$. It was known that in cosmological Einstein
gravity coupled to a massive vector, the Lifshitz exponents are
determined by the mass parameter. In Einstein-Weyl gravity, the mass
of the massive spin-2 mode is determined by $\alpha \Lambda$, and it
is therefore not surprising that it is also responsible for
determining the Lifshitz exponents. In general there are two allowed
values of $z$ for a given $\alpha\Lambda$.  For the
non-supersymmetric solution (\ref{ewlif}), the two values coalesce
when $\alpha\Lambda=3/2$, the critical point. The resulting
coalesced value is $z=1$ and hence there is no Lifshitz solution in
critical Einstein-Weyl pure gravity \cite{lppp}. The situation is
quite different for supersymmetric Lifshitz solutions in
Einstein-Weyl supergravity. The allowed values for $z$ are given by
\begin{equation}
\alpha \Lambda = \fft{(z+2)^2}{6z}\qquad \longrightarrow \qquad z=-2
+ 3\alpha \Lambda \pm \sqrt{3\alpha\Lambda (3\alpha \Lambda -4)}\,.
\end{equation}
For $z$ to be real, we must have $\alpha \Lambda \le 0$ or $\alpha
\Lambda \ge 4$. At the critical point $\alpha \Lambda = 3/2$, the
allowed values for $z$ are 1 and 4.  Thus in addition to the AdS$_4$
vacuum, critical supergravity allows $z=4$ Lifshitz solution as
well, in which the massive vector $A_\1$ is turned on.  The two
values of $z$ coalesce when $\alpha \Lambda= 4/3$ for which $z=2$.

     It should be remarked that our solutions demonstrate explicitly
that the auxiliary fields in off-shell supergravities can play a
non-trivial role in constructing supersymmetric solutions once they
develop dynamics after higher-order invariants are involved.

\section{Supersymmetric asymptotic Lifshitz and AdS solutions}

It was shown in \cite{lppp} that Einstein-Weyl pure gravity not only
admits Lifshitz vacuum solutions (\ref{ewlif}), but also black holes
that are asymptotic to those Lifshitz vacua.  However, owing to the
complexity of the non-linear differential equations, no known
examples of exact asymptotic Lifshitz solutions were constructed
except in the conformal gravity limit.  In this section, we consider
supersymmetric asymptotic Lifshitz and also AdS solutions with the
plane isometry. The existence of supersymmetry reduces the equations
of motion drastically, and enables us to obtain exact solutions in
some special cases.  For large number of solutions, we still have to
use the numerical approach even if the equations are simplified
significantly.

Let us start with the more general membrane-like ansatz
\begin{equation}
ds^2 = \fft{dr^2}{f} - a dt^2 + r^2 (dx^2 + dy^2)\,,\qquad A_\1 =
c\, dt\,,\label{genans}
\end{equation}
where the functions $a, f$ and $c$ depend on the coordinate $r$
only. This ansatz contains the Lifshitz solutions (\ref{lifs-ans})
after some trivial scaling of the coordinates $(x,y)$. The vielbein
and the corresponding spin connection are chosen to be
\begin{eqnarray}
&&e^{\hat 0}=\sqrt{a}\, dt\,,\qquad e^{\hat x} = r
dx\,,\qquad e^{\hat y}=r dy\,,\qquad e^{\hat r} =
\fft{dr}{\sqrt{f}}\,.\cr 
&&\omega^{\hat 0}{}_{\hat r} = \fft{a'\sqrt{f}}{2a}\,,
\qquad \omega^{\hat x}{}_{\hat r}=\fft{\sqrt{f}}{r}
e^{\hat x}\,,\qquad \omega^{\hat y}{}_{\hat
r}=\fft{\sqrt{f}}{r} e^{\hat y}\,,
\end{eqnarray}
Following the analogous analysis of the Killing spinor equation in
the previous section, we find that the existence of a Killing spinor
implies that
\begin{equation}
\fft{\sqrt{f}\,a'}{4a} - \fft{c}{3\sqrt{a}} - \ft16 S=0\,,\qquad
\fft{\sqrt{f}}{2r} + \fft{c}{6\sqrt{a}} - \ft16 S=0\,,\label{k0fo}
\end{equation}
where $S$ is given by (\ref{scalareom}). Thus we find that
\begin{equation}
c=\fft{3\sqrt{a}\Big(\sqrt{-\Lambda/3}\, r -
\sqrt{f}\Big)}{r}\,,\qquad
\fft{a'}{a}=\fft{2\sqrt{-3\Lambda}}{\sqrt{f}} -
\fft{4}{r}\,.\label{aceom}
\end{equation}
Substituting these into the bosonic equations of motion, we find
that equations are reduced to the following second-order non-linear
differential equation for $f$:
\begin{equation}
\alpha \Big(3\sqrt{f}(r f'' - 5f' + 4\Lambda r) + 5\sqrt{-3\Lambda}
(r f' + 2f)\Big) - 2r (\sqrt{-3\Lambda}\, r -
3\sqrt{f})=0\,.\label{f2peom}
\end{equation}
It is clear that if we turn off the vector field, namely by setting
$c=0$, we have only the AdS vacuum solution, given by
\begin{equation}
f=-\ft13\Lambda\, r^2\,,\qquad a=r^2\,.
\end{equation}
It is also straightforward to verify that the Lifshitz solutions
obtained in the previous section satisfy these equations.  An
important question to ask is whether an extremal black hole solution
can emerge in this system.  The answer is no for non-vanishing
$\Lambda$ and finite $\alpha$.  This is because if $a$ has a zero at
$r=r_0>0$, a necessary requirement for a black hole, it follows from
the second equation of (\ref{aceom}) that the function $f$ must have
a double root at $r_0$.  However, it follows from (\ref{f2peom})
that such a double root for $f$ is not possible for non-vanishing
$\Lambda$. In what follows we shall present some exact solutions in
special cases and numerical solutions for general parameters.

\subsection{Exact solutions in Einstein-Weyl supergravity}

If one is able to solve for $f$ in (\ref{f2peom}), one can obtain
$a$ and $c$ straightforwardly from (\ref{aceom}). We have not obtained
the general solution for (\ref{f2peom}). However, for some special
parameters, we are able to obtain exact solutions.

  Let us set $\Lambda=0$, in which case, we find that (\ref{f2peom})
can be solved completely, giving the following local solution.
\begin{eqnarray}
ds^2&=&4\alpha \Big(\fft{dr^2}{r^2-r_0^2 + c\, r^6} -r^{-4} dt^2 +
r^2 (dx^2 + dy^2)\Big)\,,\cr 
A_\1 &=&- \fft3{r^3}\sqrt{r^2-r_0^2+c\, r^6}\,dt\,.
\end{eqnarray}
In presenting the solution, we have scaled the coordinates $t, x$
and $y$ appropriately.  The Riemann-square of the metric is given by
\begin{equation}
\alpha^2\, {\rm Riem}^2 =\ft{27}4 + \ft34 c\, r^2 (9c\, r^6+10r^2-6
r_0^2) -\fft{3r_0^2(22r^2-15r_0^2)}{4r^4}\,.
\end{equation}
Thus we see that for vanishing $c$ and $r_0$, the solution is the
Lifshitz with $z=-2$ and the metric is regular everywhere. For $c=0$
but $r_0\ne 0$, the solution has a curvature singularity at $r=0$.
For $r_0=0$ but $c\ne 0$, the singularity is located at $\infty$.
For both non-vanishing parameters, the solution has a curvature
singularity at both $r=0$ and $\infty$.  The singularity at
$r=\infty$ for non-vanishing $c$ is a null singularity, since it
takes an infinity physical time for the light geodesic to reach the
singularity:
\begin{equation}
t \sim \sqrt{c} \int^\infty \fft{dr}{r} \rightarrow \infty\,.
\end{equation}

First let us consider the case with $c=0$. The solution describes a
smooth wormhole that connects two symmetric Lifshitz worlds with
$z=-2$. The curvature singularity at $r=0$ is geodesically
disconnected from the wormhole. To see this more clearly, it is
instructive to write the metric in proper radial coordinate, namely
\begin{equation}
ds^2=4\alpha \Big(d\rho^2 - (\cosh\rho)^{-4} dt^2 + (\cosh\rho)^2
(dx^2 + dy^2)\Big)\,,\qquad A_\1=\fft{3\sinh\rho}{(\cosh\rho)^3}\,
dt\,.\label{exactwh}
\end{equation}
The solution requires that $\alpha >0$, and this implies that the
vector field is a ghost.  If on the other hand $\alpha<0$, we can
perform appropriate analytic continuation $(x,y)\rightarrow {\rm i}
(x,y)$ and treat the coordinate $\rho$ as the time variable. We
obtain the BPS cosmological solution
\begin{equation}
ds^2=4(-\alpha) \Big(-dt^2 + (\cosh t)^{-4} dz^2 + (\cosh t)^2 (dx^2
+ dy^2)\Big)\,,\qquad A_\1=\fft{3\sinh t}{(\cosh t)^3} dz\,.
\end{equation}

    Now let us consider the solution with $r_0=0$, for which it is
advantageous to use $\rho=1/r$ as the radial coordinate.  We have
\begin{eqnarray}
ds^2 &=& 4\alpha \Big(\fft{d\rho^2}{\rho^2 W} - \rho^4 dt^2 +
\fft{1}{\rho^2} (dx^2 + dy^2)\Big)\,,\qquad A_\1 = -3
\rho^2\sqrt{W}\, dt\,,\cr 
W&=& 1- \fft{a}{\rho^4}\,.
\end{eqnarray}
The solution are asymptotic to the $z=-2$ Lifshitz solution in the
UV region.  For $a>0$, the solution is a wormhole that connects to
the two asymptotic regions.  for $a<0$, the solution can be viewed
as Lifshitz black hole, with a null singularity at $\rho=0$. Since
the horizon and the singularity coincides, the temperature is zero,
consistent with supersymmetry.  Furthermore, the vector field $A_\1$
is finite and non-vanishing at the horizon $\rho=0$.

\subsection{Exact solutions in conformal supergravity}

    For constant $S$ given in (\ref{scalareom}), the equations for
BPS solutions are given by (\ref{f2peom}) with $\alpha$ sent to
infinity.  We find an exact solution for $\alpha\rightarrow \infty$,
given by
\begin{eqnarray}
ds^2&=&-\fft{(r-r_0)^6}{r^4} dt^2 + \fft{dr^2}{(r-r_0)^2} + r^2
(dx^2 + dy^2)\,,\cr A_\1 &=& \fft{3r_0(r-r_0)^3}{r^3}dt\,,\qquad
S=3\,.\label{confsol1}
\end{eqnarray}
In conformal supergravity, the vector is massless and this solution
describes a charged extremal black brane in conformal supergravity,
with the electric charge per unit area given by
\begin{equation}
Q=\alpha \int F_\2 = 9\alpha r_0^2\,.
\end{equation}
Note that the field $S$ does not appear in the bosonic action in
conformal gravity, but only in the off-shell supersymmetric
transformation rules. This solution describes an extremal BPS black
brane, with the coordinate $r$ running from the AdS$_2\times T^2$ at
the horizon $r=r_0$ to the $r\rightarrow \infty$ asymptotic AdS$_4$.

As discussed previously, in conformal supergravity, the scalar $S$
is unrestricted, not even has to be a constant.  It follows that in
(\ref{k0fo}) $S$ is a function of $r$ rather than a parameter given
in (\ref{scalareom}). For asymptotic AdS solution, we can take $a=f$
without loss of generality in conformal gravity. This leads to the
following BPS solution
\begin{eqnarray}
ds^2 &=& -f dt^2 + \fft{dr^2}{f} + r^2 (dx^2 + dy^2)\,,\qquad f=r^2
+ c_1 r + c_0\,,\cr 
A_\1 &=& - \fft{c_0}{r}dt\,,\qquad S=\fft{4c_0 + 5c_1 r+ 6
r^2}{2r\sqrt{f}}\,.\label{worm0}
\end{eqnarray}
The Riemann-square of the metric is given by
\begin{equation}
{\rm Riem}^2 = 24 + \fft{24 c_1}{r} + \fft{8 c_1 r_0}{r^3} + \fft{4
r_0^2}{r^4} + \fft{8 (c_1^2 + r_0)}{r^2}\,.
\end{equation}
Thus the curvature singularity is located at $r=0$.  This solution
describes a black hole with non-vanishing temperature when the
function $f$ has a single root. This clearly contradicts with the
fact that non-vanishing temperature breaks supersymmetry. It turns
out that when $f$ has a single root, the function $S$ is divergent
on the horizon, and hence the Killing spinors are not well defined
there. Thus although the solution has locally-defined Killing spinor
outside the horizon, the solution is not supersymmetric. When $f$
has a double root, corresponding to $f=(r-r_0)^2$, we have
\begin{eqnarray}
ds^2 &=& -(r-r_0)^2 dt^2 + \fft{dr^2}{(r-r_0)^2} + r^2 (dx^2 +
dy^2)\,,\cr 
A_\1 &=& - \fft{r_0^2}{r}dt\,,\qquad S=3 -
\fft{2r_0}{r}\,.\label{confsol2}
\end{eqnarray}
Note that in this case, the function $S$ is finite and non-vanishing
on and outside the horizon. Thus we have obtained another extremal
BPS black brane in the off-shell conformal supergravity. The
solution runs from the AdS$_2\times T^2$ at the horizon $r=r_0$ to
the AdS$_4$ at the asymptotic $r\rightarrow \infty$.  In fact it is
easy to demonstrate that this solution is conformally equivalent to
the extremal AdS black hole (\ref{confsol1}).  To see this, we can
multiply the metric in (\ref{confsol1}) by a conformal factor
$\rho^2/r^2$ with
\begin{equation}
\fft{(r-r_0)^3}{r^3} = \fft{\rho-\rho_0}{\rho}\,,
\end{equation}
and $\rho_0=3r_0$.  The resulting metric, if expressed in terms of
$\rho$ variable, is identical to that in (\ref{confsol2}).  The
conformal factor is finite and non-vanishing outside and on the
horizon and hence the two metrics have the same topology. (Examples
of conformally locally equivalent black holes with different global
structures in conformal gravity can be found in \cite{lppp}.)

We also obtain asymptotic Lifshitz solution, given by
\begin{eqnarray}
ds^2 &=& -h\, dt^2 + \fft{dr^2}{r^2\, h} + r^2 (dx^2 +
dy^2)\,,\qquad h=c_0 r^2 + c_1 + \fft{c_2}{r^2}\,,\cr 
A_\1 &=& \fft{2 c_2}{r^2}\,,\qquad S = \fft{3c_0 r^3+ 2 c_1 r^2 +
c_2}{r^2\sqrt{h}}\,.
\end{eqnarray}
The BPS condition requires that $S$ being finite and hence $h$
cannot have a single root to give rise to a black hole with
temperature.  When $S$ has a double zero at $r=r_0$, neither $c_0$
nor $c_2$ vanishes, and consequently the solution has power-law
singularities at both $r=0$ and $r=\infty$. For $c_0=0$, with $c_1$
and $c_2$ both positive, the solution is asymptotic to the Lifshitz
solution with $z=0$, but it suffers from having a naked power-law
curvature singularity at $r=0$.  If instead $c_2=0$, the the
singularity occurs at $r \rightarrow \infty$.

As mentioned in the introduction, the gauge invariance of the
bosonic action of conformal supergravity is broken by the fermionic
sector. Thus although adding a pure gauge to $A_\1$ does not affect
the bosonic equations of motion for any of the solutions discussed
in this subsection, it will break the supersymmetry.

\subsection{Numerical solutions}

Although we have obtained some exact solutions for (\ref{f2peom}),
it is not clear to us how to solve it in general. In this
subsection, we use numerical analysis to obtain asymptotic Lifshitz
and AdS solutions for generic $\alpha\Lambda$. As we have explained
earlier, there can be no extremal black hole solutions for
non-vanishing $\Lambda$. The only smooth solutions are wormholes
that connect to two AdS or Lifshitz vacua. Analogous to
(\ref{worm0}) with $c=0$, the solutions are characterized by the
following behavior of the metric functions $a$ and $f$. Running from
$r\rightarrow \infty$, the function $f$ runs from $\sim r^2$ to
encounter a single zero at the middle $r=r_0$, throughout which $a$
runs from $r^{2z}$ to a constant without being singular or
vanishing.

     To demonstrate that such a solution indeed exists for generic
$\alpha \Lambda$, we first perform the Taylor expansion in the
middle where the wormhole throat resides.  Then we perform the
Taylor expansion at the asymptotic infinity.  We treat the middle
expansion as the boundary condition and use the numerical analysis
to show that the asymptotic infinity can indeed be reached.

\subsubsection{Middle expansion}

Let us assume that the function $f$ has a single largest root at
$r=r_0$ between $r_0$ to $\infty$, we can perform Taylor expansion
around $r=r_0$, given by
\begin{equation}
f=f_1 (r-r_0) + f_2 (r-r_0)^{3/2} + f_3 (r-r_0)^2 + f_4
(r-r_0)^{5/2} + \cdots\,.
\end{equation}
We find that the low-lying coefficients are
\begin{eqnarray}
f_2&=&\fft{4\lambda(2r_0-5\alpha f_1)}{3\alpha \sqrt{f_1}}\,,\qquad
f_3= -\fft{3f_2\sqrt{f_1}}{8\lambda r_0} +
\fft{2\lambda^2(34\alpha^2 f_1^2 - 5 \alpha f_1 r_0 - 2
r_0^2)}{3\alpha^2 f_1^2}\,,\cr 
f_4 &=& -\fft{\lambda}{45\alpha^3 f_1^{7/2} r_0}\Big(3 \alpha f_1^2
(315 \alpha^2 f_1^2 - 108 \alpha f_1 r_0 - 4 r_0^2)\cr 
&&\qquad\qquad+4 \lambda^2 r_0 (260 \alpha^3 f_1^3 + 93 \alpha^2
f_1^2 r_0 - 20 r_0^3)\Big)\,,
\end{eqnarray}
where we have let $\Lambda=-3\lambda^2$.  The corresponding $a$ is
then given by
\begin{eqnarray} \fft{a}{a_0}&=&1 + a_1 \sqrt{r-r_0} +
a_2 (r-r_0) + a_3 (r-r_0)^{3/2} + \cdots\,,\cr 
a_1 &=& \fft{12\lambda}{\sqrt{f_1}}\,,\qquad a_2=-\fft{4 (\alpha
f_1^2 - 23 \alpha f_1 \lambda^2 r_0 + 2 \lambda^2 r_0^2)}{\alpha
f_1^2 r_0}\,.
\end{eqnarray}
The function $c$ is given by
\begin{equation}
c=3\lambda r_0 + \fft{3(f_1-6\lambda^2r_0)}{\sqrt{f_1}}\,
\sqrt{r-r_0} + \cdots\,.
\end{equation}
Thus we see that $r=r_0$ is not a horizon, but a wormhole throat.
Note that for the solution to have a double zero, we must have
$\lambda=0$, {\it i.e.} the cosmological constant $\Lambda$
vanishes, for which exact solutions were obtained in section 4.1. To
investigate the wormhole throat in some detail, we consider the
co-moving coordinate $\rho$, defined by $d\rho = dr/\sqrt{f}$,
namely
\begin{equation}
r=r_0 + \ft14 f_1 \rho^2 + \fft{2r_0-5\alpha f_1}{12\alpha} \rho^3 +
\cdots\,,
\end{equation}
in the middle region of $r=r_0$, where have set $\lambda=1$.  Thus
we have $\rho\rightarrow 0$ when $r$ approaches $r_0$.  For the
above expansion, we see that $r$ has the minimum $r_0$ at $\rho=0$
provided that $f_1>0$. This implies that the scale factor $r^2$ of
the $T^2$ metric $dx^2 + dy^2$ has a local minimum. The metric
function associated with $g_{tt}$ in terms of $\rho$ in the middle
is given by
\begin{equation}
\fft{a}{a_0}=1 + 6\rho + (18 - \fft{f_1}{r_0}) \rho^2 + \cdots\,.
\label{aexpansion}
\end{equation}
Thus we see that $a$ at $\rho=0$ is not the minimum.  However, as we
discussed in the preamble in section 4, $a$ cannot be zero, and it
follows that $a$ must have the minimum $a_{\rm min}>0$ for some
different $\rho$.  We shall use numerical analysis to demonstrate
that these middle expansions can be smoothly extended to the
asymptotic infinities.  Thus we establish that $r=r_0$ is not only a
local but also the global minimum, and hence our solutions in
general describe asymmetric wormholes where the minima for $r^2$ and
$a^2$ are located at different points of the radial coordinate. If
$\lambda=0$, {\it i.e.}~the cosmological constant vanishes, the
second term in (\ref{aexpansion}) vanishes and the wormhole,
discussed in section 4.1, is symmetric.

The half integer powers in the near horizon expansion cannot be
removed for non-vanishing $\Lambda$.  If we choose $\alpha$
appropriately such that $f_2=0$, it is easy to verify that the
coefficient $f_4$ is non-vanishing.  Such structures of the middle
expansions suggest that the function $f$ is intrinsically a
complicated function and cannot be expressed in terms of a rational
polynomial.

\subsubsection{Asymptotic expansion}

For any given $\alpha \Lambda$, there is always an AdS vacuum
solution. In addition, there can be two BPS Lifshitz solutions.  In
this subsection, we study the asymptotic expansions around the AdS
and Lifshitz vacua.  For simplicity, we shall set $\Lambda = -3$.
Asymptotically, we have $f=r^2/\ell^2$.  Substituting this into
(\ref{f2peom}) we find
\begin{equation}
\ell=1\,,\qquad {\rm or}\qquad \ell= \ft13(z+2)\,.\label{ellvalue}
\end{equation}
The first case gives rise to the AdS$_4$ whilst the latter case
gives rise to the Lifshitz vacuum with exponent $z$.

{\bf Asymptotic AdS expansion:}   The AdS vacuum solution exists for
all values of the $\alpha$ parameter.  We find that the large $r$
expansions around the asymptotic AdS for various functions are given
by
\begin{eqnarray}
a &=& r^2 + c_+ r^{\fft12(1+n)} + c_- r^{\fft12 (1-n)} +
\cdots\,,\cr 
f &=& r^2 + \ft16 (3-n) c_+r^{\fft12(1+n)} +\ft16 (3+n) c_-
r^{\fft12 (1-n)}+\cdots\,,\cr 
c&=&-\ft14 (3-n) c_+ r^{-\fft12(1-n)} - \ft14 (3+n) c_-
r^{-\fft12(1+n)}+ \cdots\,,
\end{eqnarray}
where $n=\sqrt{(\alpha-4)/\alpha}$. Note that when $0<\alpha<4$, the
power $n$ is complex implying some oscillatory behavior in the next
leading order.  The fact that the sub-leading term in the asymptotic
region expansion involve generic non-integer powers suggests that a
exact solution for generic $\alpha$ is unlikely. At the critical
point $\alpha=-1/2$, there is a more elegant AdS expansion, given by
\begin{eqnarray}
a&=&r^2 - \fft{2M}{r} - \fft{2M^2}{9 r^4} - \fft{469 M^3}{729 r^7} +
\cdots\,,\cr 
f &=& r^2 - \fft{2M}{r} - \fft{13M^2}{9 r^4} - \fft{469 M^3}{243
r^7} \cdots\,,\cr 
c &=& \fft{3M}{r^2} + \fft{2M^2}{3r^5} + \fft{86 M^3}{81
r^8}+\cdots\,,
\end{eqnarray}
Thus we see that there is no supersymmetric logarithmic solution at
the critical points.  In contrast, non-supersymmetric Logarithmic
black holes were known to exist in critical gravity \cite{lppp}. Our
result suggests that the supersymmetric perturbation of the AdS$_4$
vacuum is ghost free.

{\bf Asymptotic Lifshitz expansion:}  For Lifshitz solutions, the
Lifshitz exponents $z$ depends on $\alpha$.  Since we have set
$\Lambda=-3$, it follows that we have
\begin{equation}
\alpha=-\fft{(z+2)^2}{18z}\,.
\end{equation}
The expansion up to the sub-leading terms are given by
\begin{eqnarray}
a &=& \ft19 (z+2)^2 \Big[r^{2z} + c_+ r^{-\fft23 + \fft76z +
\fft16m} + c_- r^{-\fft23 + \fft76z - \fft16m}+ \cdots\Big]
\,,\cr
f&=&\fft{9}{(z+2)^2}\Big[ r^2 - \fft{1}{6(z+2)} \Big((m-4-5z) c_+
r^{\fft43 + \fft{m}6 - \fft{5z}{6}}\cr &&\qquad - (m+4+5z) c_-
r^{\fft43 - \fft{m}6 - \fft{5z}{6}}\Big)+ \cdots\Big]\,,\cr 
c&=& (z-1) r^z + \fft{1}{196(z+2)}\Big(
 (-520 + 49 m + 2 m^2 - 83 z) c_+ r^{-\fft23 +\fft{m}6 +
 \fft{z}6}\cr 
&&\qquad + (-520 - 49 m + 2 m^2 - 83 z) c_- r^{-\fft23 - \fft{m}6 +
\fft{z}6}\Big)+ \cdots\,,
\end{eqnarray}
where $m=\sqrt{64 - 32 z + 49 z^2}$.  It is worth noting that $m$ is
real for all real exponents $z$; however, there are very few
examples of $m$ being integers for integer exponents.  This suggests
that exact solutions for generic $z$ is highly unlikely.

     At the critical point, there is a Lifshitz solution with $z=4$;
however, the sub-leading term involve irrational powers since
$m=12\sqrt5$.  The best behavior occurs at $\alpha = -4/9$, where
the two Lifshitz exponents coalesce to become $z=2$.  Even in this
case, the expansion powers are not of integers but rational numbers:
\begin{eqnarray}
a &=& \ft{16}9\Big( r^4 -\fft{M}{r^{2/3}} - \fft{13M^2}{84r^{16/3}}
- \fft{185 M^3}{1176 r^{10}} + \cdots\Big)\,,\cr 
f &=& \ft9{16} \Big( r^2  - \fft{7M}{6r^{8/3}} - \fft{73M^2}{144
r^{22/3}} - \fft{1147 M^3}{3024 r^{12}} + \cdots\Big)\,,\cr 
c&=& r^2 + \fft{5M}{4r^{8/3}} + \fft{65 M^2}{336 r^{22/3}} +
\fft{1325 M^3}{9408 r^{12}} + \cdots\,.
\end{eqnarray}

\subsubsection{Numerical results}

Having obtained both the middle and asymptotic expansions, we can
use numerical analysis to verify that the middle expansion of the
wormhole throat can indeed smoothly extend to the asymptotic
infinity.  We have checked this with general parameters.  As
concrete examples, we consider two cases.  The first is the critical
supergravity, corresponding to $\alpha=-\ft12$.  If we let $r_0=10$,
we find that smooth solutions exist for a continuous parameter range
$\delta_-\le f_1 \le \delta_+$, with $(\delta_-, \delta_+)\sim
(49.98, 81.63)$. The solution becomes singular outside this range.
When $f_1=\delta_-$, the solution is asymptotic to the $z=4$
Lifshitz vacuum. For $\delta_-< f_1 \le \delta_+$, the solution
becomes asymptotic AdS.  This behavior is very similar to the AdS
and Lifshitz black holes obtained in \cite{lppp} and the smooth
solutions are all asymptotic AdS except for one which is asymptotic
to the Lifshitz vacuum.

     Another example is $\alpha=-4/9$, corresponding to having a
$z=2$ Lifshitz solution.  For $r_0=10$, we again find smooth
solutions in the parameter region $\delta_-\le f_1 \le \delta_+$,
with $(\delta_-, \delta_+)\sim (67.82, 83.00)$.  However, in this
case, the asymptotic AdS solution occurs only when $f_1=\delta_+$.
In the remaining parameter space $\delta_-\le f_1 < \delta_+$, the
solutions are asymptotic to the $z=2$ Lifshitz vacuum.

     Thus in general, for a given choice of $r_0$, we have smooth
solutions in the parameter region $\delta_-\le f_1 \le \delta_+$,
for some appropriate $(\delta_-, \delta_+)$.  Depending on the value
of $\alpha$, the solutions can be either most asymptotic AdS or most
asymptotic Lifshitz.   It should be pointed out that in the
numerical calculation, we can distinguish whether the solution is
asymptotic AdS or Lifshitz by evaluating
\begin{equation}
\ell^2 = \lim_{r\rightarrow \infty} \fft{r^2}{f}\,,
\end{equation}
and there can only be two discrete values given in (\ref{ellvalue}).
We find that in general the solutions approach these two values very
slowly with the logarithmic correction in $r$.

\section{Scaling Noether charge and supersymmetry}

It was shown in the context cosmological gravity with a massive
vector that asymptotic Lifshitz solutions have an additional
conserved Noether charge \cite{peet}.  This Noether charge was also
found in such a solution in Einstein-Weyl pure gravity \cite{lppp}.
In this section, we study the Noether charge in the context of
Einstein-Weyl supergravity and how the supersymmetry restricts the
value of this charge.

Following the procedure of \cite{peet} and also \cite{lppp}, we
consider the following ansatz
\begin{equation}
ds^2 =  - \tilde a^2 dt^2 + d\rho^2 + \tilde b^2 (dx^2 +
dy^2)\,,\qquad A_\1=\tilde c\, dt\,.
\end{equation}
It is straightforward to see that the Lagrangian is invariant under
the global scaling symmetry
\begin{equation}
\tilde a \rightarrow \lambda^2 \tilde a\,,\qquad \tilde b\rightarrow
\lambda^{-1} \tilde b\,,\qquad \tilde c\rightarrow \lambda \tilde
c\,.
\end{equation}
The corresponding conserved Noether charge can be obtained
straightforwardly.  In terms of the coordinate system
(\ref{genans}), the Noether charge is given by
\begin{eqnarray}
N &=& \fft{r\sqrt{f} (2a-ra')}{\sqrt{a}} + \fft{\alpha \sqrt{f}}{6r
\sqrt{a^5}}\Big(8 r^3 a^2 c c' + 6 a^2 f r^3 a''' -3 a^2 r^2 (2 a -
ra')f''\cr 
&& +a r^2 (10 a f - 14 f r a' + 9arf')a'' + 8 a^3f - 14 r a^2 a'f +
6 r^3 a'^3 f \cr 
&&+ 2 r a^3  f' + 4 a^2 r^2 a'f' -7 a r^3 a'^2 f'\Big)\,.
\end{eqnarray}
Substituting the supersymmetry conditions (\ref{aceom}) into the
Noether charge, we find that
\begin{equation}
N=\sqrt{a} \Big[\alpha \Big(3\sqrt{f}(r f'' - 5f' + 4\Lambda r) +
5\sqrt{-3\Lambda} (r f' + 2f)\Big) - 2r (\sqrt{-3\Lambda}\, r -
3\sqrt{f})\Big]\,,
\end{equation}
which vanishes identically because of the equation of motion
(\ref{f2peom}).  Thus we see that the supersymmetry constrains the
Noether charge to zero.  In contrast, for non-supersymmetric
Lifshitz black holes obtained in \cite{peet} and \cite{lppp}, the
Noether charge is non-vanishing and it is constant proportional to
the product of the temperature and entropy of the black hole
\cite{peet,lppp}.

\section{Integrability condition}

One important characteristics of off-shell supergravity is that the
supersymmetric transformation rules form closed super algebra
without requiring the equations of motion. This implies that the
integrability conditions for the Killing spinor equations in
off-shell supergravities have some special properties. In both
on-shell or off-shell supergravities, the integrability conditions
of Killing spinor equations are not immediately related to the
equations of motion, but rather they are related to the Riemann
tensor of the background.  However, in on-shell supergravities, a
certain $\Gamma$-matrix projected integrability condition can be
shown to vanish identically on shell.  This property no longer holds
in off-shell supergravity.  This is because the supersymmetric
transformation rules are fixed in off-shell supergravity whilst the
equations of motion can be altered by the introduction of additional
super invariants.  Thus it is worthwhile to study the
$\Gamma$-matrix projected integrability condition for the Killing
spinor equations. We do this by consider the generic background, and
then substitute the supersymmetric Lifshitz solutions to demonstrate
that the integrability conditions are indeed satisfied.  The purpose
of this exercise is not to check whether the solutions we have
obtained are indeed supersymmetric, which has already been
established.  Rather it is to highlight the distinguishing features
of off-shell supergravities.

The Killing spinor equation is given by
\begin{eqnarray}
\delta\psi_\mu = - D_\mu\epsilon + \ft{1}{6} (2 A_\mu +
  \Gamma_{\rho\mu} A^\rho)\Gamma_5 \epsilon -
 \ft16 \Gamma_\mu(S+ i\Gamma_5 P)\epsilon=0
 \end{eqnarray}
By computing $\Gamma^{\nu}D_{[\nu}D_{\mu]}$ on $\epsilon$, we have
\begin{eqnarray}
0\!\!\!\!&=&\!\!\!\!R_{\mu\nu}\Gamma^{\nu}\epsilon- \ft{2}{3}
\Gamma^{\nu}(2\nabla_{[\nu} A_{\mu]} +
  \Gamma_{\rho[\mu} \nabla_{\nu]}A^\rho)\Gamma_5 \epsilon +
\ft23 \Gamma^{\nu}\Gamma_{[\mu}(\nabla_{\nu]}S+ i\Gamma_5
\nabla_{\nu]}P)\epsilon \cr\!\!\!\!&&\!\!\!\!-\ft1{9}
\Gamma^\nu\left[ (2 A_{[\mu} +
   \Gamma_{\rho[\mu} A^\rho)\Gamma_5  -
 \Gamma_{[\mu}(S+ i\Gamma_5 P)\right]
 \left[ (2 A_{\nu]} -
  \Gamma_{\nu]\sigma} A^\sigma)\Gamma_5  -
 \Gamma_{\nu]}(S+ {\rm i}\Gamma_5 P)\right]\epsilon
\cr \!\!\!\!&=&\!\!\!\!\Big[R_{\mu\nu}+\ft2{9}  (A_\mu A_\nu -A_\rho
A^\rho g_{\mu\nu})
+\ft13(S^2+P^2)g_{\mu\nu}\Big]\Gamma^{\nu}\epsilon
\cr\!\!\!\!&&\!\!\!\! -\ft{2}{3} \Big[\ft12F_{\nu\mu} +
\nabla_{(\mu}A_{\nu)}+ \ft12 \nabla_{\rho}A^\rho
 g_{\mu\nu}\Big]\Gamma^{\nu}\Gamma_5 \epsilon
 -\ft{1}{6} F_{\nu}{}^\rho\Gamma^{\nu}{}_{\rho\mu}\Gamma_5 \epsilon
 \cr\!\!\!\!&&\!\!\!\!
+\ft13 \Gamma^{\nu}{}_{\mu}\Big[\nabla_{\nu}-\ft{2} {3}
A_\nu\Gamma_5\Big]
 (S+ {\rm i} P\Gamma_5)\epsilon
-  \Big[\nabla_{\mu}+\ft{2} {3} A_\mu\Gamma_5\Big]
 (S+ {\rm i} P\Gamma_5)\epsilon\,.\label{1gaproj}
 \end{eqnarray}
For the leading-order supergravity with $\alpha=0$, the massive
vector $A_\1$ vanishes, and it is then straightforward to see that
the integrability condition is satisfied by the equations of motion,
without further checking. However, when the $\alpha$ term is
involved, $A_\1$ becomes dynamical and the equations of motion
involve the variation of the Weyl-squared terms.  It is thus clear
that the above expression in the right will not vanish identically
on-shell. For the solution (\ref{susylif}), we find
\begin{eqnarray}
0 \!\!\!&=&\!\!\!\ell^{-1}\ft23(z-1)(z+2)r^{z}\Gamma^{\hat
0}\epsilon -\ft{2} {3}\ell^{-1} (z-1)(z+2)r^z\Gamma_5 \epsilon \cr
0\!\!\!&=&\!\!\!\ft1{9}(5z+4)(z-1)\ell^{-1}r\Gamma^{\hat i}\epsilon
-\ft{1}{3} z(z-1)\ell^{-1}r\Gamma^{\hat r\hat 0} {}_{\hat i}\Gamma_5
\epsilon - \ft{2} {9} (z-1)(z+2)\ell^{-1}r\Gamma_5
 \Gamma^{\hat 0}{}_{\hat i}\epsilon
 \cr
0\!\!\! &=&\!\!\!-\ft4{9}  (z-1)^2
\ell^{-1} r^{-1}\Gamma^{\hat r}\epsilon +\ft{2}{3}
z(z-1)\ell^{-1}r^{-1}\Gamma^{\hat 0}\Gamma_5 \epsilon -\ft{2} {9}
(z-1)(z+2)\ell^{-1}r^{-1}\Gamma_5 \Gamma^{\hat 0}{}_{\hat
r}\epsilon\,.
\end{eqnarray}
It implies that
\begin{equation}
\Gamma_{\hat r} \epsilon_0+\epsilon_0=0\,,\qquad \Gamma^{\hat
0}\Gamma_5\epsilon_0 +\epsilon_0=0\,,
\end{equation}
which are precisely the condition we have obtained in section 3.
Thus we see explicitly that the $\Gamma$-matrix projected
integrability conditions do not vanish identically by the
equations of motion, but they are subject to further constraints.

Even for the leading-order theory with $\alpha=0$, the structures of
the $\Gamma$-matrix projected integrability condition
(\ref{1gaproj}) appear rather ``unruly'' and at the first sight, it
is hard to see how such structures can cancel the supersymmetric
variation of the bosonic Lagrangian.  For example, although the
first line of the right hand side of the equation (\ref{1gaproj}) is
indeed related to the Einstein equations of motion for
$\alpha=0=\Lambda$, the remaining terms appear to be rather
disorganized. To investigate the supersymmetric invariance of the
Lagrangian in some detail, we present the the fermionic Lagrangian
up to and including the quadratic order:
\begin{equation}
e^{-1} {\cal L}_F=\bar \psi_\mu \Gamma^{\mu\nu\rho} \psi_{\nu\rho} -
2\sqrt{-\Lambda/3}\, \bar \psi_\mu \Gamma^{\mu\nu} \psi_\nu\ -
\ft23\alpha \Big(\bar \psi_{\mu\nu} \Gamma^\rho D_\rho \psi_{\mu\nu}
-\bar \psi_{\mu\lambda} \Gamma^{\mu\nu\rho} D_\rho
\psi_\nu{}^\lambda + \cdots\Big)\,,
\end{equation}
where $\psi_{\mu\nu} = 2 D_{[\mu} \psi_{\nu]}$ and the ellipses
inside the last bracket denote the quadratic fermion terms of the
type $\psi^2 \times \nabla$(bosonic fields).  Note that for
$\alpha=0$, the auxiliary fields do not appear at all in the
quadratic fermionic action.  The supersymmetric transformation rule
for the gravitino was given in (\ref{susytrans}) and the ones for
the bosonic fields are \cite{lppp}
\begin{eqnarray}
\delta e_{\mu}{}^a &=& \bar \epsilon \Gamma^a \psi_\mu\,,\cr 
\delta S &=& \ft12 \bar \epsilon \Gamma^{\mu\nu} \widehat
\psi_{\mu\nu}\cr &=&\ft12\bar\epsilon
\left[\Gamma^{\mu\nu}\psi_{\mu\nu} + A^\nu \Gamma_5\psi_{\nu} - (S+
i\Gamma_5 P)\Gamma^{\nu} \psi_{\nu}\right]\,,\cr 
\delta P&=&\ft12{\rm i}\bar \epsilon \Gamma^{\mu\nu} \Gamma_5
\widehat \psi_{\mu\nu}\cr 
&=&  \ft12{\rm i}\bar\epsilon
\Gamma_5\left[\Gamma^{\mu\nu}\psi_{\mu\nu} +  A^\nu
\Gamma_5\psi_{\nu} - (S+ {\rm i}\Gamma_5 P)\Gamma^{\nu}
\psi_{\nu}\right]\,,\cr 
\delta A_{\mu} &=&\ft18 \bar \epsilon (\Gamma_\mu \Gamma^{\nu\rho} -
3 \Gamma^{\nu\rho} \Gamma_\mu) \Gamma_5 \widetilde \psi_{\mu\nu}\cr
&=&\bar\epsilon(\delta_{\mu}^{[\nu}\Gamma^{\rho]}-\ft14\Gamma_\mu
{}^{\nu\rho})\Gamma_5\psi_{\nu\rho} +\bar\epsilon\left(-\ft12
A_{\mu}\Gamma^{\rho}+\delta_{\mu}^{\rho}A_{\nu}\Gamma^{\nu}
-\ft14A^\sigma\Gamma^{\rho}{}_{\sigma\mu}\right) \psi_{\rho}\cr &&
-\ft12(S\Gamma_5+ {\rm i} P)\psi_{\mu} \,,\label{susy1}
\end{eqnarray}
where
\begin{eqnarray}
\widehat \psi_{\mu\nu} &=& \psi_{\mu\nu} +\ft13 \Gamma_5 (2 A_{[\mu}
+ A^\rho \Gamma_{\rho[\mu})\psi_{\nu]} + \ft13\Gamma_{[\mu}(S + {\rm
i} \Gamma_5 P)\psi_{\nu]}\,,\cr 
\widetilde\psi_{\mu\nu} &=&\psi_{\mu\nu} -\ft{1}{3} \Gamma_5 \left(4
A_{[\mu} - A^\rho \Gamma_{\rho[\mu} \right) \psi_{\nu]}-\ft13
\Gamma_{[\mu}(S+ i\Gamma_5 P)\psi_{\nu]}\,.
\end{eqnarray}
Thus we see that the supersymmetric transformation rule of the
bosonic field in the on-shell supermultiplet, namely the vielbein,
is standard and it does not involve a derivative of gravitino. This
is a generic feature for on-shell supergravities, and it is
responsible for the familiar structure of the first line in the
right hand side of equation (\ref{1gaproj}). On the other hand, the
supersymmetric transformation rules for the auxiliary fields $S,P$
and $A_\mu$ all involve the gravitino field strength, a derivative
of the gravitino. This implies that the mechanism for the
supersymmetric invariance of the Lagrangian in off-shell
supergravities is rather different from that in on-shell
supergravities.  To elaborate this further, let us consider the
following
\begin{eqnarray}
-\Gamma^{\mu\nu\rho}D_{\nu}\delta\psi_\rho &=& \ft12
\Big[R^{\mu}{}_{\rho}-\ft12\delta^{\mu}_{\rho}R+\ft1{9}  (2A^\mu
A_\rho +A_\lambda A^\lambda \delta^{\mu}_{\rho})
-\ft13(S^2+P^2)\delta^{\mu}_{\rho} \Big]\Gamma^{\rho}\epsilon \cr
&&-\ft{1}{12} F_{\nu\rho} \Gamma^{\mu\nu\rho}\Gamma_5 \epsilon
-\ft{1}{3} \Big[\nabla_{(\nu}A_{\rho)}-\nabla_{\lambda}A^\lambda
g_{\nu\rho}-\ft{1}{2} F^{\mu}{}_{\rho}\Big]\Gamma^{\rho}\Gamma_5
\epsilon \cr
&& +\ft1{3} \Gamma^{\mu\rho}\Big[\nabla_{\rho}+\ft{1} {3}
A_\rho\Gamma_5\Big] (S+ i P\Gamma_5)\epsilon -\ft{1} {3}
A^\mu\Gamma_5 (S+ {\rm i} P\Gamma_5)\epsilon \cr 
&&-\ft1{6}
\left[(\Gamma^{\mu\nu\rho}A_{\rho}+4A^{[\mu}\Gamma^{\nu]})\Gamma_5 -
2\Gamma^{\mu\nu}(S+ {\rm i}\Gamma_5 P)\right]
\delta\psi_{\nu}\,.\label{3gammaproj}
\end{eqnarray}
In on-shell supergravities, the above quantities involving the
manifest $\epsilon$ terms would precisely cancel the variation of
the bosonic Lagrangian whilst the $\delta\psi_\nu$ term would cancel
the variation of the Yukawa terms in supergravities.\footnote{For
this reason, it was shown that a bosonic gravity system that has
consistent Killing spinor equations can be promoted to
pseudo-supergravity whose action, up to and including quadratic
order in fermions, is invariant under the pseudo-supersymmetry
\cite{Lu:2011zx}-\cite{Liu:2011ve}. If the field contents coincide
with a known supermultiplet, the theory becomes indeed true
supergravity. On the other hand, if the field contents do not form a
supermultiplet, the pseudo-supersymmetric transformation rules will
not form a known (finite-dimensional) super-Poincar\'e algebra, and
the corresponding theory is called pseudo-supergravity
\cite{Lu:2011zx}-\cite{Liu:2011ve}.} This clearly is not true for
(\ref{3gammaproj}), even though the first term in the right hand
size of  (\ref{3gammaproj}) indeed cancels the variation of bosonic
Lagrangian with $\alpha=0=\Lambda$.  It is thus worth checking
explicitly how the Lagrangian is invariant under the supersymmetry.

    The Lagrangian for Einstein-Weyl supergravity we have considered
in this paper in fact contains three independent invariants
\cite{lppp}
\begin{eqnarray}
{\cal L} &=& {\cal L}^\1 + 2\sqrt{-\Lambda/3}\,{\cal L}^\2 +
\ft12\alpha {\cal L}^\3\,,\cr 
{\cal L}^\1 &=& {\cal L}^\1_B + {\cal L}^\1_F =\sqrt{-g} \Big(R -
\ft23 (A_\1^2 + S^2 + P^2)\Big) + \sqrt{-g} \bar \psi_\mu
\Gamma^{\mu\nu\rho} \psi_{\nu\rho}\,,\cr 
{\cal L}^\2 &=& {\cal L}^\2_B + {\cal L}^\2_F =2\sqrt{-g}\, S -
\sqrt{-g}\, \bar\psi_\mu \Gamma^{\mu\nu} \psi_\nu\,,\cr 
{\cal L}^\3 &=& {\cal L}^\3_B + {\cal L}^\3_F = \sqrt{-g}
\Big(C^{\mu\nu\rho\sigma} C_{\mu\nu\rho\sigma} + \ft13
F_\2^2\Big)\,,\cr 
&&\qquad\qquad\quad -\ft13 \sqrt{-g} \Big(\bar \psi_{\mu\nu}
\Gamma^\rho D_\rho \psi_{\mu\nu} -\bar \psi_{\mu\lambda}
\Gamma^{\mu\nu\rho} D_\rho \psi_\nu{}^\lambda+ \cdots\Big)\,.
\end{eqnarray}
For simplicity and the purpose of demonstrating the distinguishing
feature of off-shell supergravity, we shall present only the
supersymmetric invariance of the action associated with ${\cal
L}^\1$ and ${\cal L}^\2$ up to and including the quadratic order in
fermion.  We find
\begin{eqnarray}
&&\int d^4x \delta(\sqrt{-g}\,{\cal L}^{(1)}_B)\cr 
&&= \int d^4x\,\sqrt{-g} \Big( -2\left[R^{\mu \rho}-\ft23A^\mu
A^\rho-\ft12g^{\mu\rho}\left(R -\ft23(A^2 +
S^2+P^2)\right)\right]\bar\epsilon\Gamma_{\rho}\psi_\mu \cr 
&&\qquad\qquad\qquad- \ft{4}{3}\bar\epsilon(A^{[\mu}\Gamma^{\nu]}-
\ft14A_{\rho}\Gamma^{\rho\mu\nu})\Gamma_5 \widetilde\psi_{\mu\nu}
-\ft23\bar\epsilon\Gamma^{\mu\nu}(S+ {\rm i}\Gamma_5 P)
\widehat\psi_{\mu\nu}\Big)\,,\cr 
&&\int d^4x \delta(\sqrt{-g}\,{\cal L}^{(1)}_F)\cr 
&&= \int d^4x\, \sqrt {-g} \Big(
2\left(R^{\mu}{}_{\rho}-\ft12\delta^{\mu}_{\rho}R\right)
\bar\epsilon\Gamma^{\rho}\psi_\mu +\ft{4}{3}\bar\epsilon\left(A^\mu
\Gamma^{\nu}-\ft14A_\rho\Gamma^{\mu\nu\rho}\right)
\Gamma_5\psi_{\mu\nu} \cr 
&&\qquad\qquad\qquad +\ft23\bar\epsilon\Gamma^{\mu\nu}(S+ {\rm
i}\Gamma_5 P)\psi_{\mu\nu}\Big]\Big)\,,\cr 
&&\int d^4x \delta(\sqrt{-g}\,{\cal L}^{(2)}_B)=\int d^4x\,
\sqrt{-g} \Big( 2S\,\bar\epsilon\Gamma^{\mu}
\psi_\mu+\bar\epsilon\Gamma^{\mu\nu}
\widehat\psi_{\mu\nu}\Big)\,,\cr 
&& \int d^4x \delta(\sqrt{-g}\,{\cal L}^{(2)}_F)=\int d^4x\,
\sqrt{-g} \,\Big (-\bar\epsilon \Gamma^{\mu\nu}\psi_{\mu\nu} -\bar
\epsilon \Big[A^\mu\Gamma_5 + \Gamma^{\mu}(S-{\rm i}\Gamma_5
P)\Big]\psi_\mu\Big)\,.
\end{eqnarray}
In the above, we have performed appropriate integration by parts and
dropped the surface terms.  It is now a straightforward calculation
to verify that the each action associated with ${\cal L}^\1$ and
${\cal L}^\2$ is independently invariant under the supersymmetry. It
becomes clear that the cancelation of the last term in
(\ref{3gammaproj}) is not by the Yukawa term, as would be the case
in on-shell supergravity, but due to the gravitino field strength
that appears in the transformation rules of the auxiliary fields
$S,P$ and $A_\1$.

\section{Spherically-symmetric solutions}

The asymptotic AdS and Lifshitz solutions we have obtained so far
are characterized by the flat level surfaces and the metric is brane
like.  It is natural to investigate whether there are
spherically-symmetric solutions that are also supersymmetric.
Spherically-symmetric AdS and Lifshitz-like black holes were shown
to exist in both Einstein-Weyl and conformal gravities \cite{lppp}.
Supersymmetric and static asymptotic AdS solutions are common
occurrence in gauged supergravities, although they typically suffer
from having a naked power-law curvature singularity. In this
section, we study the supersymmetry of spherically-symmetric ansatz
and demonstrate that the only supersymmetric such a solution is the
AdS$_4$ vacuum. There are no asymptotic AdS or Lifshitz-like
solutions that are both spherically-symmetric and supersymmetric in
Einstein-Weyl or conformal supergravities.

Let us consider the ansatz:
\begin{equation}
ds^2 = \fft{dr^2}{f} - a dt^2 + r^2 (d\theta^2 + \sin^2\theta
d\phi^2)\,,\qquad A_\1 = c dt\,,
\end{equation}
where the functions $a,f$ and $c$ depend on $r$ only. A natural
choice for the vielbein is
\begin{equation}
e^{\hat 0}=\sqrt{a}\, dt\,,\qquad e^{\hat 1} = r d\theta\,,\qquad
e^{\hat 2}=r \sin\theta d\phi\,,\qquad e^{\hat r} =
\fft{dr}{\sqrt{f}}\,.
\end{equation}
The corresponding spin connection has the following non-vanishing
components
\begin{equation}
\omega^{\hat 0}{}_{\hat r} = \fft{a'\sqrt{f}}{2a}e^{\hat 0}\,,
\qquad \omega^{\hat 1}{}_{\hat r}=\fft{\sqrt{f}}{r} e^{\hat
1}\,,\qquad \omega^{\hat 2}{}_{\hat r}=\fft{\sqrt{f}}{r} e^{\hat
2}\,, \qquad \omega^{\hat 2}{}_{\hat
1}=\fft{\cos\theta}{r\sin\theta} e^{\hat 2}\,,
\end{equation}
We decompose the $\Gamma$-matrices as follows
\begin{eqnarray}
\Gamma_{\hat 0}=i\sigma_{1}\otimes \sigma_3\,,~~~\Gamma_{\hat
r}=\sigma_{2}\otimes \sigma_3\,,~~~\Gamma_{\hat a}= 1 \otimes
\sigma_{\hat a}\,,~~~ \Gamma_5=i\Gamma^{\hat0\hat1\hat2\hat
r}=\sigma_{3}\otimes \sigma_{3}\,,
\end{eqnarray}
where $\sigma_i$ are Pauli matrices. We now perform the reduction on
$S^2$ by introducing two component spinor $\chi_{\pm}$ which obey
the equations
\begin{eqnarray}\label{SpinorOnS2}
{\nabla}'_a \chi_\pm=\pm\frac{\rm i}{2}{\sigma}_{\hat a}\chi_\pm\,,
\end{eqnarray}
where ${\nabla}'$ is defined with the spin connection on the unit
sphere. It is related to the covariant derivative on the spinor in
the sphere direction of our ansatz as
\begin{eqnarray} \label{spincon}
\nabla_a={\nabla'}_a+\frac{1}{4}\Omega_{abr}\Gamma^{br}=
{\nabla'}_a+\frac{\sqrt f}{4r}\Gamma_a{}^{\hat r},
\end{eqnarray}
Covariance under $SU(2)$ transformations ensures that we can take
\begin{eqnarray}
\chi_-={\rm i}\sigma_3\chi_+
\end{eqnarray}
without loss of generality. Now we can expand the spinor $\epsilon$
over basis on the sphere
\begin{eqnarray}\label{DecompSpinXi}
\epsilon=\sum_{\alpha=+,-}\epsilon_{\alpha}\otimes\chi_{\alpha} .
\end{eqnarray}
The Killing spinor equation in the $t$, $r$ and the sphere
directions implies that
\begin{eqnarray}
0&=&\sum_{\alpha}\Big[\partial_{r} + \fft{c}{6\sqrt a}
\Gamma_{r}{}^{\hat 0} \Gamma_5 \epsilon + \ft16\Gamma_r S \Big]
\epsilon_{\alpha}\otimes\chi_{\alpha}\cr 
&=&\sum_{\alpha}\Big[\partial_{r}\epsilon_{\alpha}\otimes\chi_{\alpha}
+\fft{c}{6\sqrt {af}} \epsilon_{\alpha}\otimes\sigma_3\chi_{\alpha}
+ \fft1{6\sqrt f} S
\sigma_2\epsilon_{\alpha}\otimes\sigma_3\chi_{\alpha}\Big]\,,\cr
0&=& \sum_{\alpha}\Big[\partial_{t} +\fft{a'\sqrt
f}{4a}\Gamma_{t}{}^{\hat r} - \fft{c}{3} \Gamma_5 \epsilon +
\ft16\Gamma_t S \Big] \epsilon_{\alpha}\otimes\chi_{\alpha}
\cr&=&\sum_{\alpha}\Big[\partial_{t}\epsilon_{\alpha}\otimes
\chi_{\alpha}+\fft{a'\sqrt f}{4\sqrt
a}\sigma_3\epsilon_{\alpha}\otimes\chi_{\alpha}- \fft{c}{3}
\sigma_3\epsilon_{\alpha}\otimes\sigma_3\chi_{\alpha} - \fft{i\sqrt
a}{6} S
\sigma_1\epsilon_{\alpha}\otimes\sigma_3\chi_{\alpha}\Big]\,,\cr
0&=& \sum_{\alpha}\Big[\nabla_{a}' +\frac{\sqrt
f}{2r}\Gamma_a{}^{\hat r} + \fft{c}{6\sqrt a} \Gamma_{a}{}^{\hat 0}
\Gamma_5 \epsilon + \ft16\Gamma_a S \Big]
\epsilon_{\alpha}\otimes\chi_{\alpha} \cr&=&\sum_{\alpha}e^{\hat
a}_a\Big[\frac{i}{2r}\alpha\epsilon_{\alpha}\otimes{\sigma}_{\hat
a}\chi_{\alpha} +\frac{\sqrt
f}{2r}\sigma_2\epsilon_{\alpha}\otimes\sigma_{\hat
a}\sigma_3\chi_{\alpha}\cr
&&\qquad\qquad -\fft{c}{6\sqrt
a}\sigma_2\epsilon_{\alpha}\otimes\sigma_{\hat a}\chi_{\alpha} +
\ft16 S \epsilon_{\alpha}\otimes\sigma_{\hat a}\chi_{\alpha}\Big]\,,
\end{eqnarray}
where $S=3\sqrt{-\Lambda/3}$. It follows that
\begin{eqnarray}
&&\partial_{r}\epsilon_{+} +\fft{{\rm i}\,c}{6\sqrt {af}}
\epsilon_{-} + \fft{\rm i}{6\sqrt {f}} S \sigma_2\epsilon_{-}=0\,,
\cr&&\partial_{r}\epsilon_{-} -\fft{{\rm i}\,c}{6\sqrt {af}}
\epsilon_{+}- \fft{\rm i}{6\sqrt {f}} S \sigma_2\epsilon_{+}=0\,,
\cr&& 
\partial_{t}\epsilon_{+}+\fft{a'\sqrt f}{4\sqrt a}\sigma_3\epsilon_{+}
-\fft{{\rm i}\,c}{3} \sigma_3\epsilon_{-}+ \fft{\sqrt a}{6} S
\sigma_1\epsilon_{-}=0\,, \cr&& 
\partial_{t}\epsilon_{-}+\fft{a'\sqrt f}{4\sqrt a}\sigma_3
\epsilon_{-}+ \fft{{\rm i}\,c}{3} \sigma_3\epsilon_{+} - \fft{\sqrt
a}{6} S \sigma_1\epsilon_{+}=0\,, \cr&& 
\frac{\rm i}{2r}\epsilon_{+} +\frac{{\rm i}\sqrt
f}{2r}\sigma_2\epsilon_{-} -\fft{c}{6\sqrt a}\sigma_2\epsilon_{+} +
\ft16 S \epsilon_{+}=0\,, \cr 
&& -\frac{\rm i}{2r}\epsilon_{-} -\frac{{\rm i}\sqrt
f}{2r}\sigma_2\epsilon_{+} -\fft{c}{6\sqrt a}\sigma_2\epsilon_{-} +
\ft16 S \epsilon_{-}=0\,, \label{ksres}
\end{eqnarray}
It is clear that the Killing spinors can be expressed as
$\epsilon_\pm =e^{{\rm i} E_\pm t} \tilde \epsilon_\pm (r)$. It
follows that the last four equations of (\ref{ksres}) are
effectively eight linear equations for the four variables $\tilde
\epsilon_\pm^i$ with $i=1,2$.  For $c=0$, we find that a solution
exists, giving rise to precisely the AdS$_4$ vacuum with the
cosmological constant $\Lambda$. When $c$ is non-vanishing, we find
that the Killing spinor equations simply has no solution before even
checking whether the background satisfies the bosonic equations of
motion.

     It is rather intriguing that Einstein-Weyl or conformal
off-shell supergravities favor the existence of AdS and Lifshitz
solutions with the plane isometry rather than the
spherically-symmetric solutions. In usual supergravities, the
opposite is true: the number of BPS spherically-symmetric solutions
is far greater than that of the supersymmetric Lifshitz solutions
which typically involve non-trivial hypermultiplets.

\section{Conclusions}

In this paper, we have studied Einstein-Weyl supergravity and
conformal supergravity.  The supersymmetry are realized in the
off-shell formalism.  The field content of the theory consists of
the metric, a massive vector and one complex scalar as the bosonic
fields and a gravitino field as their supersymmetric partner.  In
the off-shell formalism, the closure of the supersymmetry
transformation rules is realized without needing the input of the
equations of motion. This implies that one can alter the Lagrangian
and hence the equations of motion by adding new super invariants
without changing the supersymmetry.  It follows that in detail
calculation, the supersymmetric invariance of the action is realized
very differently than that in the usual on-shell supergravities.  In
particular, the $\Gamma$-matrix projected integrability condition
for the Killing spinor equations does not vanish identically on
shell. Ostensibly, the supersymmetric invariance of the action is
unlikely to be true.  The invariance is saved by the fact that the
supersymmetric transformation rules for the auxiliary fields, namely
the massive vector and the complex scalar, involve derivative of the
gravitino. We present the explicit demonstration of the invariance
up to and including the quadratic order of fermions in the action.

The main purpose of the paper is to construct supersymmetric
solutions in Einstein-Weyl and conformal supergravities.  We find
that in addition to the previously-known supersymmetric AdS vacuum,
the theory admits a large class of BPS Lifshitz vacua supported by
the massive vector.  The Lifshitz exponents $z$ are solely
determined by the product of the cosmological constant and the
coupling of the Weyl-squared term.  We demonstrate by obtaining
explicit Killing spinors that these solutions preserve $\fft14$ of
the supersymmetry.  The result is intriguing in that the massive
vector field belongs to the auxiliary fields in the off-shell
supermultiplet.  For the lowest-order supergravity, these fields are
set to zero by the equations of motion.  Thus one might have
expected that these fields play no role in physics.  However, when
higher-derivative super invariants are added into the Lagrangian,
the auxiliary fields can pick up kinetic terms and become dynamical.
Our solutions are the first examples that not only these auxiliary
fields in off-shell supergravity can be used to construct
non-trivial solutions, they also play an essential role for the
solutions to be supersymmetric.  Since the auxiliary fields are not
related to the central charge of the super algebra, the
supersymmetry of our solutions is not realized by the standard BPS
condition that balances the energy and the central charge. Instead,
it is likely the case that the massive spin-2 hair balances the
massive vector hair such that the Noether charge associated with the
scaling symmetry of the plane-symmetric ansatz vanishes. Thus our
solutions demonstrate a new mechanism of achieving supersymmetry of
a bosonic background in higher-derivative off-shell supergravities.

We then obtain the equations of backgrounds that are supersymmetric
and asymptotic to the AdS and Lifshitz solutions. Although the
theory involve fourth-order derivatives, the equations of motion
constrained by the supersymmetric are significantly simpler and
reduced to one second-order non-linear differential equation. For
some special choice of parameters, we have obtained many examples of
exact solutions, including asymptotic Lifshitz solutions with $z=-2$
and extremal AdS black holes. We then use numerical analysis to
study the structure of general solutions. We find that for
non-vanishing cosmological constant, the general solutions are
wormholes that are asymptotic to Lifshitz vacua.  For a given size
of the wormhole throat, we find that there exists a parameter region
in which the solutions are smooth.  Depending on the value of the
product of the cosmological constant and the coupling of the
Weyl-squared term, the smooth wormholes are either mostly asymptotic
AdS with the asymptotic Lifshitz solution lying at one boundary of
the parameter space, or the vice versa. We also study the
supersymmetric condition of spherically-symmetric solutions and we
find that the only solution is the AdS vacuum.

Higher-derivative gravity in general suffers from having ghost modes
in its spectrum. Einstein-Weyl supergravity is not different
\cite{lppp}. However, such modes are not disastrous in condensed
matter physics. Moreover, the supersymmetry that our solutions
possess might help to stabilize the solutions. For
example, we have demonstrated that the ghost log mode in critical
supergravity does not contribute to the supersymmetric asymptotic
AdS solution. The large number of supersymmetric solutions that are
asymptotic to the AdS and Lifshitz vacua we have constructed in this
paper demonstrate that higher-derivative off-shell supergravities
can have important applications in the area of the AdS/CFT
correspondence.

\section*{Acknowledgement}

H.L.~would like to thank the organizer and the participants of the
advanced workshop ¡°Dark Energy and Fundamental Theory¡±, supported
by the Special Fund for Theoretical Physics from the National
Natural Foundation of China for useful discussions. The research of
H.L.~is supported in part by NSFC grant 11175269.

\end{document}